\documentclass[conference]{IEEEtran}
\IEEEoverridecommandlockouts
\usepackage{cite}
\usepackage{amsmath,amssymb,amsfonts}
\usepackage{algorithmic}
\usepackage{graphicx}
\usepackage{textcomp}
\usepackage{xcolor}
\usepackage{kantlipsum}
\usepackage{fancyhdr}
\usepackage{lastpage}
\def\BibTeX{{\rm B\kern-.05em{\sc i\kern-.025em b}\kern-.08em
    T\kern-.1667em\lower.7ex\hbox{E}\kern-.125emX}}
\begin{document}
\begin{titlepage}
    \begin{center}
        \vspace*{1cm}
            
        \Huge
        \textbf{Exploiting and Securing Docker Containers and Kubernetes Pods from a MitM attack}

            
        \vspace{0.5cm}
        \LARGE
        An Academic research paper for possible submission to an IEEE conference 
            
        \vspace{1.5cm}
            
        \textbf{Henry Kabuye}
            
        \vfill
        Submitted in partial requirements for the degree of\\
        Msc Cybersecurity
            
        \vspace{0.8cm}
            
        \includegraphics[width=0.4\textwidth]{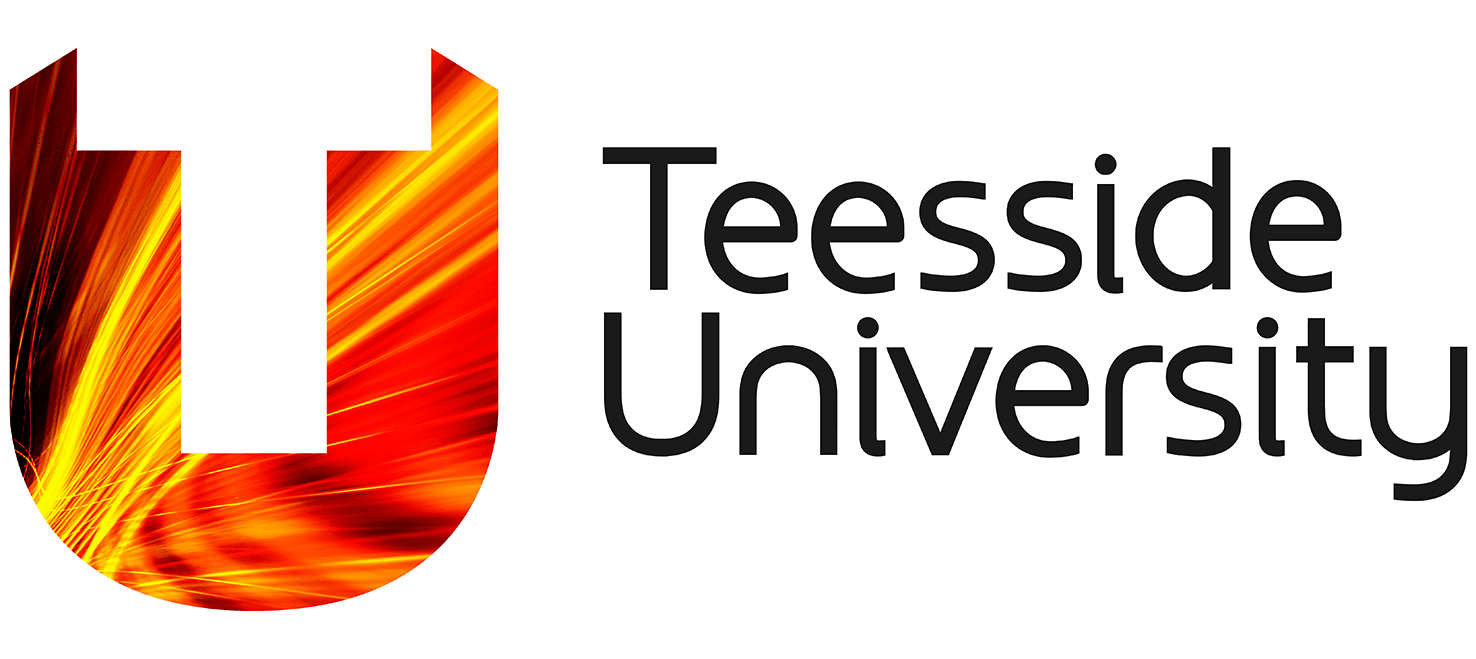}
            
        \Large
        School of Computing, Engineering and Digital Technologies\\
        United Kingdom\\
        
        January 11, 2022

    \end{center}
    Supervisor: Ismail Khalid Kazmi\\
    2nd Reader: Chunyan Mu\\
\end{titlepage}

\title{Exploiting and Securing Docker containers and Kubernetes pods from a MitM attack\\
\thanks{Teesside University, Middlesbrough, TS1 3BA, UK}
}

\author{\IEEEauthorblockN{Henry Kabuye}
\IEEEauthorblockA{\textit{School of Computing,} \\\textit{Engineering \& Digital Technologies} \\
\textit{Teesside University} UK} \\
\and
\IEEEauthorblockN{Ismail Khalid Kazmi}
\IEEEauthorblockA{\textit{School of Computing,} \\\textit{Engineering \& Digital Technologies} \\
\textit{Teesside University} UK} \\
\and
\IEEEauthorblockN{Chunyan Mu}
\IEEEauthorblockA{\textit{School of Computing,} \\\textit{Engineering \& Digital Technologies} \\
\textit{Teesside University} UK }\\
\and
\IEEEauthorblockN{Paolo Modesti}
\IEEEauthorblockA{\textit{School of Computing,} \\\textit{Engineering \& Digital Technologies} \\
\textit{Teesside University} UK }\\
}

\maketitle

\begin{abstract}

\emph{PURPOSE -} Workloads in containers, such as Docker containers and Kubernetes pods, are just as vulnerable as workloads in non-container environments. Hackers may use many of their older techniques, like the phishing exploits, application exploits and network intrusions on these containers. This article is a systematic review research project based on design and creation that explores several novel techniques for effectively securing containerised-based operating systems from "Man in the Middle" (MitM) attack. The use of a conceptual model associated with the level of abstraction utilised for the top-level representation of communication and cryptographic primitives is a unique feature of the proposed framework in this article. The security mechanisms also include the use of a security library named "AnBxJ java library" and container firewalls based on a layer 7 of the OSI (Open System Interconnect) model. Answers to the research question; \emph{"How can containerised-based operating systems be effectively secured from Man-in-the-Middle attack?"}, is found in this article. This article aims to assist practitioners in protecting their containerised pod installations by systematising data about Docker security practices and providing a ‘zero trust’ architecture on such containerised-based operating systems.

\emph{METHODOLOGY -} The research methodology used in this analysis is a Systematic Review (SR) composed of the Preferred Reporting Items for Systematic Reviews and Meta-Analyses (PRISMA). PRISMA relates to a compilation of results of evidence-based reviews that are in line with the same researched topic leading to evidence-based solutions. 

\emph{FINDINGS -} Success factors were identified, and a security mechanism was implemented successfully on a containerised-based operating system scenario.

\emph{VALUE -} The discovered, according to the research question, may be beneficial for practitioners in protecting their Kubernetes and Docker container installations by systematising information about Kubernetes security practices and providing a ‘zero trust’ architecture on their containerised-based operating systems.

\end{abstract}


\section{Introduction}  Containers, such as Docker, have influenced the development and deployment of many software systems\cite{Reis2022docker, bhardwaj2021virtualization}. Containers are virtualization abstractions at the operating system level for executing separated systems on a host with a single kernel \cite{Casalicchio2019Container9}. A container image is a small, standalone executable package that contains all the necessary components to execute cloud applications, such as system libraries, code, settings and system tools \cite {Casalicchio2019Container9}. To safeguard digital assets and minimise the attack surface accessible to the adversary, it is critical to design resilient and secure systems \cite{Modesti2020Formal,laschi2008design,oates2022researching}. The author proposed a security mechanism in conjunction with a security library "AnbxJ java library" to provide cryptographic primitives and an API for communication between the containerised-based platforms. The proposed mechanism was based on the AnB (Alice and Bob) notation to provide verification of abstract and concrete models as well as automatic generation of Java implementation \cite{Modesti2020Formal}. Through the proposed mechanisms of creating a secure architecture that was deployed on a containerised-based platform scenario, cyber attacks such as "Man-in-the-middle (MitM)" attacks were mitigated.

Several commercial cloud service providers, such as Amazon, Google, IBM, and Microsoft, are investing increasingly on micro-service deployment utilising containerized environments to enable cloud service flexibility and quick response\cite{BachiegaSBS2018Container}. OpenVZ, Docker and Kubernetes are all well-known containerization technologies. Kubernetes, often known as \emph{"K8s"}, is an open-source system used in this article for automatic deployment, scaling, and administration of applications in containers\cite{kubernetes2021}. The automated coordination, management, and configuration of computer systems, applications, and services is known as \emph{orchestration} \cite{kubernetes2021}. According to Corodescu \emph{et al} \cite{CorodescuNKSMPR2021Locality}, complicated processes and workflows in Information technology (IT) are managed more easily with orchestration.

A zero-trust mechanism is deployed as the cybersecurity paradigm based on resource protection and the assumption that implicit trust should not be granted and then often evaluated \cite{rose2020zero,goodman2021zerotrust,Modesti2020Formal}. Studies \cite{oates2022researching,Modesti2020Formal}, have shown that cryptography is critical for secure authentication on containerised-based enable communication systems. Rather than filtering traffic based on IP addresses, the author proposed a layer 7 firewall based on a java security library to provide an API with cryptographic primitives that can examine the contents of data packets to see whether they include malware or other cyber threats\cite{Modesti2020Formal}. 
The proposed architecture is set up on a virtualised laboratory platform called the vSphere VMware platform provided by Teesside University. Virtual laboratories can play a critical role in addressing the aforesaid difficulties in containerised-based settings, particularly in the context of this project and the ethical considerations, where combining conceptual knowledge with technological abilities gains strategic relevance\cite{laschi2008design}. A virtual laboratory was setup to provide compelling options to solve the concerns mentioned above.

In the remaining sections of this article: Section II is the review of literature; section III are the methods used in the research findings; section IV are the results of the research findings; section V is the discussion of the research findings; section VI demonstrates the threats model and sections VII-IX, demonstrates a proposed security mechanism, implementation and evaluation of securing containerised-based operating systems.

\textbf{Research Question (RQ)}
"How can containerised-based operating systems be effectively secured from Man-in-the-Middle exploitation?"

\subsection {Rationale of the Study}
Hackers are more likely to use many of their older techniques, like the phishing exploits, application exploits and network intrusions on containerised-based operating systems. According to the Common Vulnerabilities and Exposures (CVE) \cite{mitre2021cve25737}, a security vulnerability \emph{"CVE-2021-25737"} in Kubernetes was uncovered, allowing a user to re-route pod traffic to private networks on a Node. Endpoint internet protocols (IPs) in the localhost or link-local range are already prevented by Kubernetes, however the same check was not done on Endpoint Slice IPs. As a result, revealing an external IP address to access an application in a cluster is more likely to result in a problem of a "Man-in-the-Middle (MitM)" attack. Thus, the importance of this study to effectively address this vulnerability. 
\subsection {Technical Objectives}
The main objective was to create a secure and working architecture for the Docker containers and Kubernetes pods from Man-in-the-Middle exploitation. The traffic flow between the database and the web applications has to be secured, and only authorised applications and users can access the databases. 
\subsection {Financial Objectives}
The task of securing the Docker containers and Kubernetes, strongly welcomes the costs factor since these platforms are based on cloud environments that have most security features chargeable. The "high" cost of cloud services is the source of most of the present annoyance. In this article, possible solutions to these issues were addressed. 
\subsection {Other objectives}
-To determine whether the security of Docker containers and Kubernetes can be exploited through a Man-in-the-middle attack.\\
-To identify the security mechanisms that can be deployed to protect Docker containers and Kubernetes pods from exploitation.\\
-To develop a mechanism for protecting containerised based operating systems from man-in-the-middle exploitation.\\
-To test the effectiveness of the mechanism in protecting containerised operating systems against man-in-the-middle attacks.

\section{Review of literature}
The associated work in this section does not only give an overview of secondary research but also evaluates the primary studies on the issue. This in return explains how they connect to the contributions in this article to integrate the evidence.

The necessity to build collaboration between those who develop software (Dev) and the operations team who manage production systems (Ops) created the notion of DevOps (Ops)\cite{loukides2012devopsreport, abbariAPT2016what}. Continuous Delivery (CD) and Continuous Integration (CI) pipelines are simple to manage and deploy using Docker containers and Kubernetes pods. Studies \cite{HeidariLS16QoS, Casalicchio2019Container9} show that these Docker containers and Kubernetes pods enables DevOps teams to manage container resources based on their requirements, and they even supports zero-downtime updates, rolling updates that allow incremental container updates. Studies \cite{DissanayakaMGK2020vulnerability,BachiegaSBS2018Container} demonstrate that many typical Software-defined Networking (SDN) and Transfer Layer Security (TLS) mis-configurations are caused by using the improper cypher suites.  According to TrendMicro \cite{trendmicro2020predictions} report of 2020, one of the top security concerns for development and operations teams in companies has been vulnerabilities in container components deployed with cloud architecture.

Studies \cite{CelestiMFVP2016Exploring, autio2021securing, ZhongB2020cost} show that Linux is the ideal operating system for the cloud orchestration mechanisms due to its durability, broad ecosystem, good hardware support, outstanding performance, and low cost. According to Biederman and Networx \cite{biederman2006multiple}, Linux was initially considered in 2006 and the kernel namespaces functionality required to support containers in Linux matured. This puts Linux in the best position as the baseline operating system used in the artefact during this study. Docker containers are simple to set up on cloud platforms and on-premise infrastructures \cite{Casalicchio2019Container9,kubernetes2021,BachiegaSBS2018Container,autio2021securing,ZhongB2020cost}. However, a study \cite{ahamed2021security} demonstrated that the inherent packages and libraries in the containerised images of the host kernel are the source of the majority of the vulnerabilities.

\section{Methods}
Following Preferred Reporting Items for Systematic Reviews and Meta-Analyses (PRISMA), the research methodology used in this study is a Systematic Review (SR) relating to a compilation of results of evidence-based reviews that are in line with the same researched topic leading to evidence-based solutions \cite{oates2022researching, linger2016agile, cohen2017research}. The activities of the study are conducted systematically and logically interconnected, thus none of them is a random activity or value\cite{oates2022researching}. The author independently screened out abstracts and papers for inclusion as well as carrying out data extraction and bias assessment.
\subsection {Eligibility Criteria}
The language considered is English and the years considered are between 2018 to 2022.
\subsection {Information Sources}
The digital database information sources considered are IEEE, ACM (Association for Computing Machinery) and Scopus database.
\subsection {Search Strategy}
A systematic search of literature concerning containerised based operating system and security mechanisms was performed. According to Jakobovic et al \cite{JakobovicPMW2021Toward} , Boolean functions are employed in a broad range of industries, including telecommunications, coding theory, and cryptography. The Boolean operators AND, OR, and NOT were used to limit or expand the search results. This triggered a search term; \textbf{\emph{"docker" OR "kubernetes" OR contain* AND secur*}}
\subsection {Selection Process}
The selection process was drawn on to the \textbf{"key concepts"} of the study while answering a set of questions drafted by the author to guide the research project to the RQ. 
These key concepts include:
\begin{enumerate}
  \item \texttt{Docker} containers.
  \item \texttt{Kubernetes} pods.
  \item \texttt{Security} in containerised-based operating systems.
  \item \texttt{"Man-in-the-middle (MitM)"} attack.
\end{enumerate}
In order to maximise proof recovery in the databases while avoiding needless searches, Khadjesari \emph{et al} \cite{khadjesari2011can} recommends using the \textbf{"PIO (Problem Intervention Outcome) element"} in the selection process and creation of a study question. The search strategy was made easier by populating the Boolean operators into each PIO element as illustrated in table \ref{tab1}.

\begin{table}[htbp]
\caption{PIO Element}
\begin{center}
\resizebox{\columnwidth}{!}{%
\begin{tabular}{|p{0.2\linewidth} | p{0.4\linewidth}| p{0.4\linewidth}|}
\hline
\multicolumn{3}{|c|}{\textbf{PIO Element}} \\\hline
 \textbf{Component} & \textbf{Description} & \textbf{Use of Boolean operators}\\\hline
 \textbf{\textit{P - Problem}} & MiTM attack & "Man in the middle"  OR  "MitM"  OR  "Man-in-the-Middle"\\\hline
 \textbf{\textit{I - Interventions}} & Exploiting and Securing Docker containers and Kubernetes pods & "docker" OR "kubernetes" OR contain*\\\hline
 \textbf{\textit{O - Outcome}} & Secure containerised-based operating systems & AND secur*\\\hline
\end{tabular}}
\label{tab1}
\end{center}
\end{table}

A set of four (4) questions drafted by the author guiding the research project to the RQ include:
\begin{enumerate}
  \item What has been published on \texttt{Containerised \- based} operating systems?
  \item What are the current \texttt{peer reviewed} work published on containerised-based operating systems?
  \item How do the authors handle \texttt{security} on containerised-based operating systems following the principles of security (Confidentiality, Integrity and Availability)?
  \item How can containerised-based operating systems be effectively secured from \texttt{Man-in-the-Middle} exploitation?
\end{enumerate}
\subsection {Data Collection process}
This data collection process was guided by a set of four (4) questions as mentioned in the earlier sub-section and populated in tables \ref{tab2}, \ref{tab3} and \ref{tab4} through a detailed search on ACM, IEEE and Scopus digital libraries respectively.
\begin{table}[htbp]
\caption{ACM Database}
\begin{center}
\resizebox{\columnwidth}{!}{%
\begin{tabular}{|p{0.3\linewidth} | p{0.43\linewidth}|p{0.16\linewidth} | p{0.11\linewidth}|}
\hline
\multicolumn{4}{|c|}{\textbf{ACM Digital Database}} \\\hline
 \textbf{QUESTION} & \textbf{KEYWORDS} & \textbf{FILTERS} & \textbf{No. of Articles} \\\hline
 \textbf{\textit{What has been published on Containerised-based operating systems?}} & "docker" OR "kubernetes" OR contain* AND secur* & None & 98,007\\\hline
 \textbf{\textit{What are the current peer reviewed work published on containerised-based operating systems?}} & "docker" OR "kubernetes" OR contain* AND secur* AND [Publication Date: (01/01/2018 TO 31/12/2021)] & Year: 2018 to 2022 Included: Peer reviewed. Excluded: Newspapers, Book reviews. Under: Computer Science & 3,659 \\\hline
 \textbf{\textit{How do the authors handle security on containerised-based operating systems following the principles of security (Confidentiality, Integrity and Availability)?}} & [[Title: "docker"] OR [Title: "kubernetes"] OR [[Title: contain*] AND [Title: secur*]]] AND [[Abstract: "docker"] OR [Abstract: "kubernetes"] OR [[Abstract: contain*] AND [Abstract: secur*]]] AND [[Keywords: "docker"] OR [Keywords: "kubernetes"] OR [[Keywords: contain*] AND [Keywords: secur*]]] AND [Publication Date: (01/01/2018 TO 31/12/2021)] & Year: 2018 to 2022 Included: Peer Reviewed Excluded: Newspapers, Book reviews.  Limit to: Title, Abstract and Author keywords. Discipline: Computer Science & 56 \\\hline
 \textbf{\textit{How can containerised-based operating systems be effectively secured from Man-in-the-Middle exploitation?}} & [[Title: "docker"] OR [Title: "kubernetes"] OR [[Title: contain*] AND [Title: secur*]]] AND [[Abstract: "docker"] OR [Abstract: "kubernetes"] OR [[Abstract: contain*] AND [Abstract: secur*]]] AND [[Keywords: "docker"] OR [Keywords: "kubernetes"] OR [[Keywords: contain*] AND [Keywords: secur*]]] AND [[All: "man in the middle"] OR [All: "mitm"] OR [All: "man-in-the-middle"]] AND [Publication Date: (01/01/2018 TO 31/12/2021)] & Year: 2018 to 2022 Included: Peer Reviewed Excluded: Book reviews, Newspapers Limit to: Title, Abstract and Author keywords. Discipline: Computer Science & 1 \\\hline
\end{tabular}}
\label{tab2}
\end{center}
\end{table}

\begin{table}[htbp]
\caption{IEEE Database}
\begin{center}
\resizebox{\columnwidth}{!}{%
\begin{tabular}{|p{0.3\linewidth} | p{0.43\linewidth}|p{0.16\linewidth} | p{0.11\linewidth}|}
\hline
\multicolumn{4}{|c|}{\textbf{IEEE Digital Database}} \\\hline
 \textbf{QUESTION} & \textbf{KEYWORDS} & \textbf{FILTERS} & \textbf{No. of Articles} \\\hline
 \textbf{\textit{What has been published on Containerised-based operating systems?}} & "docker" OR "kubernetes" OR contain* AND secur* & None & 11,844\\\hline
 \textbf{\textit{What are the current peer reviewed work published on containerised-based operating systems?}} & "docker" OR "kubernetes" OR contain* AND secur* AND [Publication Date: (01/01/2018 TO 31/12/2021)] & Year: 2018 to 2022 Included: Peer reviewed. Excluded: Newspapers, Book reviews. Under: Computer Science & 4,432 \\\hline
 \textbf{\textit{How do the authors handle security on containerised-based operating systems following the principles of security (Confidentiality, Integrity and Availability)?}} & ("Abstract":"docker" OR "Abstract":"kubernetes" OR "Abstract":contain* AND "Abstract":secur*) AND ("Document Title":"docker" OR "Document Title":"kubernetes" OR "Document Title":contain* AND "Document Title":secur*) AND ("Author Keywords":"docker" OR "Author Keywords":"kubernetes" OR "Author Keywords":contain* AND "Author Keywords":secur*) & Year: 2018 to 2022 Included: Peer Reviewed Excluded: Newspapers, Book reviews, TUP publisher.  Limited to: Title, Abstract and Author keywords. Discipline: Computer Science & 50 \\\hline
 \textbf{\textit{How can containerised-based operating systems be effectively secured from Man-in-the-Middle exploitation?}} & ("Document Title":"docker" OR "Document Title":"kubernetes" OR "Document Title":contain* AND "Document Title":secur*) AND ("Abstract":"docker" OR "Abstract":"kubernetes" OR "Abstract":contain* AND "Abstract":secur*) AND ("Author Keywords":"docker" OR "Author Keywords":"kubernetes" OR "Author Keywords":contain* AND "Author Keywords":secur*) AND ("Full Text \& Metadata":"Man in the middle" OR "Full Text \& Metadata":"MitM" OR "Full Text \& Metadata":"Man-in-the-Middle") & Year: 2018 to 2022 Included: Peer Reviewed Excluded: Book reviews, Newspapers Limit to: Title, Abstract and Author keywords. Discipline: Computer Science & 5 \\\hline
\end{tabular}}
\label{tab3}
\end{center}
\end{table}

\begin{table}[htbp]
\caption{Scopus Database}
\begin{center}
\resizebox{\columnwidth}{!}{%
\begin{tabular}{|p{0.3\linewidth} | p{0.43\linewidth}|p{0.16\linewidth} | p{0.11\linewidth}|}
\hline
\multicolumn{4}{|c|}{\textbf{Scopus Digital Database}} \\\hline
 \textbf{QUESTION} & \textbf{KEYWORDS} & \textbf{FILTERS} & \textbf{No. of Articles} \\\hline
 \textbf{\textit{What has been published on Containerised-based operating systems?}} & "docker" OR "kubernetes" OR contain* AND secur* & None & 197,607\\\hline
 \textbf{\textit{What are the current peer reviewed work published on containerised-based operating systems?}} & "docker" OR "kubernetes" OR contain* AND secur* AND [Publication Date: (01/01/2018 TO 31/12/2021)] & Year: 2018 to 2022 Included: Peer reviewed. Excluded: Newspapers, Book reviews. Under: Computer Science & 19,757 \\\hline
 \textbf{\textit{How do the authors handle security on containerised-based operating systems following the principles of security (Confidentiality, Integrity and Availability)?}} & ( TITLE ( "docker"  OR  "kubernetes"  OR  contain*  AND  secur* )  AND  ABS ( "docker"  OR  "kubernetes"  OR  contain*  AND  secur* )  AND  KEY ( "docker"  OR  "kubernetes"  OR  contain*  AND  secur* ) )  AND  PUBYEAR  >  2017  AND  PUBYEAR  <  2022  AND  ( LIMIT-TO ( SUBJAREA ,  "COMP" ) )  AND  ( EXCLUDE ( DOCTYPE ,  "ch" )  OR  EXCLUDE ( DOCTYPE ,  "re" ) ) & Year: 2018 to 2022 Included: Peer Reviewed Excluded: Newspapers, Book reviews.  Limited to: Title, Abstract and Author keywords. Discipline: Computer Science & 74 \\\hline
 \textbf{\textit{How can containerised-based operating systems be effectively secured from Man-in-the-Middle exploitation?}} & ( TITLE ( "docker"  OR  "kubernetes"  OR  contain*  AND  secur* )  AND  ABS ( "docker"  OR  "kubernetes"  OR  contain*  AND  secur* )  AND  KEY ( "docker"  OR  "kubernetes"  OR  contain*  AND  secur* ) )  AND  PUBYEAR  >  2017  AND  PUBYEAR  <  2022  AND  ALL ( "Man in the middle"  OR  "MitM"  OR  "Man-in-the-Middle" )  AND  ( LIMIT-TO ( SUBJAREA ,  "COMP" ) )  AND  ( EXCLUDE ( DOCTYPE ,  "ch" )  OR  EXCLUDE ( DOCTYPE ,  "re" ) ) & Year: 2018 to 2022 Included: Peer Reviewed Excluded: Book reviews, Newspapers Limit to: Title, Abstract and Author keywords. Discipline: Computer Science & 1 \\\hline
\end{tabular}}
\label{tab4}
\end{center}
\end{table}

\subsection {Data Items}
Foroughi and Luksch \cite{Faroughi2018Data} defined four (4) general phases to completing the data items of a cybersecurity project. The \textbf{first phase} is to define the problem by posing a security challenge. The security challenge in this matter was a "Man-in-the-Middle" attack and related vulnerabilities on containerized-based operating systems. 

In the \textbf{second phase}, gather required information in line with the problem statement and relevant formula. In this case, the information gathered is through the ACM, IEEE and Scopus databases that are proven to give relevant information about the issue and in line with computing subjects. 

The obtained data should be used in the \textbf{third phase}, during the analysis process, to offer appropriate data that may be used to anticipate or solve the identified problem. 

The \textbf{final phase} is a production step, which involves deploying required modules as well as a system to conduct the entire process automatically and on a regular basis when needed. And this final stage is demonstrated in a scenario artefact deployment of a java security library to automatically detect and provide cryptographic primitives for security within the containerized based operating systems. 

\section{Results}
\subsection {Study Selection related to Inclusion and Exclusion criteria}
ACM, IEEE and Scopus databases recovered an initial total of 98,007; 11,844 and 197,607 articles respectively without any filters inserted. This was meant to answer the author's question \emph{"What has been published on containerised-based operating systems?"} Considering the eligibility criteria of the author, the English language and the years between 2018 to 2022 were filtered and hence narrowing down to a total of 3,659; 4,432 and 19,757 articles for ACM, IEEE and Scopus database respectively. Including the peer reviewed articles at this stage, answers the author's second question; \emph{"What are the current peer reviewed work published on containerised-based operating systems?"} Furthermore, book reviews and newspapers were excluded from the search as well as limiting it to the \emph{"Title", "Abstract" and "Author keywords"}, which resulted into a total of 56; 50 and 74 for ACM,IEEE and Scopus database respectively. To narrow down the search results to enable them point towards the intended RQ the \emph{AND "Man in the middle"  OR  "MitM"  OR  "Man-in-the-Middle"} filter was implemented into the search strategy. And finally, a total of seven (7) articles were found relevant to the research question after inserting in the necessary filters into the search strategy with a ratio 1:5:1 for ACM, IEEE and Scopus respectively. 
\begin{table}[htbp]
\caption{Study Selection}
\begin{center}
\resizebox{\columnwidth}{!}{%
\begin{tabular}{|p{0.83\linewidth} | p{0.17\linewidth}|}
\hline
\multicolumn{2}{|c|}{\textbf{Included}} \\\hline
 \textbf{PUBLICATION} & \textbf{PUBLISHER}\\\hline
 \textbf{\textit{A. Mailewa, S. Mengel, L. Gittner and H. Khan, "Vulnerability Prioritization, Root Cause Analysis, and Mitigation of Secure Data Analytic Framework Implemented with MongoDB on Singularity Linux Containers.," ACM, 2020.}} \cite{DissanayakaMGK2020vulnerability} & ACM\\\hline
 \textbf{\textit{W. S. S. Ahamed, P. Zavarsky and B. Swar, "Security Audit of Docker Container Images in Cloud Architecture," in 2021 2nd International Conference on Secure Cyber Computing and Communications (ICSCCC), Jalandhar, 2021.}} \cite{ahamed2021security} & IEEE\\\hline
 \textbf{\textit{J. Wenhao and L. Zheng, "Vulnerability Analysis and Security Research of Docker Container," in 2020 IEEE 3rd International Conference on Information Systems and Computer Aided Education (ICISCAE), Dalian, 2020.}} \cite{wenhao2020vulnerability} & IEEE\\\hline
 \textbf{\textit{S. Sultan, I. Ahmad and T. Dimitriou, "Container Security: Issues, Challenges, and the Road Ahead," IEEE Access, vol. 7, pp. 52976 - 52996, 2019.}} \cite{SultanAD2019Container}& IEEE\\\hline
 \textbf{\textit{B. Pearson and D. Plante, "Secure Deployment of Containerized IoT Systems," in 2020 SoutheastCon, Raleigh, 2020.}} \cite{pearson2020secure}& IEEE\\\hline
 \textbf{\textit{A. Tomar , D. Jeena, P. Mishra and R. Bisht, "Docker Security: A Threat Model, Attack Taxonomy and Real-Time Attack Scenario of DoS," in 2020 , Noida, 2020.}} \cite{tomar2020docker}& IEEE\\\hline
 \textbf{\textit{W. Zhang, K. Li, T. Li, S. Niu and Z. Gao, "CM-droid: Secure container for android password misuse vulnerability," Computers, Materials and Continua, vol. 59, no. 1, pp. 181-198, 2019.}} \cite{zhang2019cm} & Scopus\\\hline
\end{tabular}}
\label{tab5}
\end{center}
\end{table}

\subsection {Results of Syntheses}
The synthesis of the PIO element of the included papers serves as a foundation for interpreting the review's conclusions and are a valuable review output in and of themselves found in tables \ref{tab6}, \ref{tab7}, \ref{tab8}, \ref{tab9}, \ref{tab10}, \ref{tab11}, \ref{tab12}. The illustrated results in tables \ref{tab6}, \ref{tab7}, \ref{tab8}, \ref{tab9}, \ref{tab10}, \ref{tab11}, \ref{tab12} are tabulated by screening out six (6) itemised fields below.
\begin{enumerate}
  \item \texttt{Publication Author} of the study selection.
  \item \texttt{Framework tools} used in the study.
  \item \texttt{Assessment tools} used in the study.
  \item \texttt{Baseline Operating System} used in the study.
  \item \texttt{Related Vulnerabilities tested} during the study.
  \item \texttt{Mitigation Techniques} suggested or used in the study.
\end{enumerate}

\begin{table}[htbp]
\caption{Result 1 of Syntheses}
\begin{center}
\resizebox{\columnwidth}{!}{%
\begin{tabular}{|p{0.3\linewidth} | p{0.7\linewidth}|}
\hline
\multicolumn{2}{|c|}{\textbf{Result 1}} \\\hline
 \textbf{Publication Author} & Mailewa et al \cite{DissanayakaMGK2020vulnerability}\\\hline
 \textbf{Framework Tools} & MongoDB \\\hline
 \textbf{Assessment Tools} & OpenVas, Dagda, PortSpider, MongoAudit, Nmap, Metasploit-Framework, Nessus \\\hline
 \textbf{Baseline Operating System} & Linux \\\hline
 \textbf{Related Vulnerabilities Tested} & TCP timestamp issues,  Log files issues,  MongoDB version exposed, ICMP timestamp issues, Network traffic restriction issues, SSL/TLS cipher issues, MongoDB port issues,  OpenSSH issues, Postfix service issues,  Blacklist and blacklist issues, DoS  issues, GRUB bootloader issues, TLS/SSL encryption issues, User authentication issues, /tmp directory issues. \\\hline
 \textbf{Mitigation Techniques} & Vulnerability Prioritization, Root causes analysis, Prevention Techniques \cite{tomar2020docker}\\\hline
\end{tabular}}
\label{tab6}
\end{center}
\end{table}

\begin{table}[htbp]
\caption{Result 2 of Syntheses}
\begin{center}
\resizebox{\columnwidth}{!}{%
\begin{tabular}{|p{0.3\linewidth} | p{0.7\linewidth}|}
\hline
\multicolumn{2}{|c|}{\textbf{Result 2}} \\\hline
 \textbf{Publication Author} & Ahamed, Zavarsky and Swar \cite{ahamed2021security}\\\hline
 \textbf{Framework Tools} & Google Docker Images \\\hline
 \textbf{Assessment Tools} & Google Container Registry \\\hline
 \textbf{Baseline Operating System} & Google Cloud (PaaS) \\\hline
 \textbf{Related Vulnerabilities Tested} & Package version issues, Default user Privilege issues, Configuration issues \\\hline
 \textbf{Mitigation Techniques} & Security Audit \\\hline
\end{tabular}}
\label{tab7}
\end{center}
\end{table}

\begin{table}[htbp]
\caption{Result 3 of Syntheses}
\begin{center}
\resizebox{\columnwidth}{!}{%
\begin{tabular}{|p{0.3\linewidth} | p{0.7\linewidth}|}
\hline
\multicolumn{2}{|c|}{\textbf{Result 3}} \\\hline
 \textbf{Publication Author} & Wenhao and Zheng \cite{wenhao2020vulnerability}\\\hline
 \textbf{Framework Tools} & Docker Schema \\\hline
 \textbf{Assessment Tools} & File system isolation, Remote isolation and communication isolation, host resource constraints and Device management, image transmission and Network isolation \\\hline
 \textbf{Baseline Operating System} & SELinux \\\hline
 \textbf{Related Vulnerabilities Tested} & Dos issues, Root privilege issues, MAC flooding issues, Man in the middle issues.\\\hline
 \textbf{Mitigation Techniques} & Linux Security Modules(LSM), AppArmor, Privileges capability mechanism, Secure Computation Mode(Seccomp), mandatory access control, Netfilter, Linux kernel Integrity protection technology, Log audit, security threat detection, Streamline operation system \\\hline
\end{tabular}}
\label{tab8}
\end{center}
\end{table}

\begin{table}[htbp]
\caption{Result 4 of Syntheses}
\begin{center}
\resizebox{\columnwidth}{!}{%
\begin{tabular}{|p{0.3\linewidth} | p{0.7\linewidth}|}
\hline
\multicolumn{2}{|c|}{\textbf{Result 4}} \\\hline
 \textbf{Publication Author} & Sultan, Ahmad and Dimitriou \cite{SultanAD2019Container}\\\hline
 \textbf{Framework Tools} & Docker\\\hline
 \textbf{Assessment Tools} & Docker Image Vulnerability Analysis Framework (DIVA) \cite{ShuGE2017docker} \\\hline
 \textbf{Baseline Operating System} & Linux \\\hline
 \textbf{Related Vulnerabilities Tested} & Meltdown and spectra attacks \\\hline
 \textbf{Mitigation Techniques} & CGroup mechanisms, namespace mechanisms, capability mechanism, Secure Computation Mode(Seccomp), Linux Security Modules(LSM) \\\hline
\end{tabular}}
\label{tab9}
\end{center}
\end{table}

\begin{table}[htbp]
\caption{Result 5 of Syntheses}
\begin{center}
\resizebox{\columnwidth}{!}{%
\begin{tabular}{|p{0.3\linewidth} | p{0.7\linewidth}|}
\hline
\multicolumn{2}{|c|}{\textbf{Result 5}} \\\hline
 \textbf{Publication Author} & Pearson and Plante \cite{pearson2020secure}\\\hline
 \textbf{Framework Tools} & Docker\\\hline
 \textbf{Assessment Tools} & N/A\\\hline
 \textbf{Baseline Operating System} & Linux \\\hline
 \textbf{Related Vulnerabilities Tested} & man in the middle attack \\\hline
 \textbf{Mitigation Techniques} & N/A \\\hline
\end{tabular}}
\label{tab10}
\end{center}
\end{table}

\begin{table}[htbp]
\caption{Result 6 of Syntheses}
\begin{center}
\resizebox{\columnwidth}{!}{%
\begin{tabular}{|p{0.3\linewidth} | p{0.7\linewidth}|}
\hline
\multicolumn{2}{|c|}{\textbf{Result 6}} \\\hline
 \textbf{Publication Author} & Tomar \emph{et al} \cite{tomar2020docker}\\\hline
 \textbf{Framework Tools} & Docker\\\hline
 \textbf{Assessment Tools} & NMap\\\hline
 \textbf{Baseline Operating System} & Linux \\\hline
 \textbf{Related Vulnerabilities Tested} & Malware, Dos attack, privilege escalation, Escape attack, ARP spoofing, MAC flooding, MitM, Unauthorised access, Poisoned images, out-dated software, Attack using CCAT, Malicious code injection, cryptojacking, Tampering, Kernel exploit \\\hline
 \textbf{Mitigation Techniques} & N/A \\\hline
\end{tabular}}
\label{tab11}
\end{center}
\end{table}

\begin{table}[htbp]
\caption{Result 7 of Syntheses}
\begin{center}
\resizebox{\columnwidth}{!}{%
\begin{tabular}{|p{0.3\linewidth} | p{0.7\linewidth}|}
\hline
\multicolumn{2}{|c|}{\textbf{Result 7}} \\\hline
 \textbf{Publication Author} & Zhang \emph{et al} \cite{zhang2019cm}\\\hline
 \textbf{Framework Tools} & Android\\\hline
 \textbf{Assessment Tools} & CM-Droid\\\hline
 \textbf{Baseline Operating System} & Linux \\\hline
 \textbf{Related Vulnerabilities Tested} & Password issues \\\hline
 \textbf{Mitigation Techniques} & CM-Droid secure container \\\hline
\end{tabular}}
\label{tab12}
\end{center}
\end{table}

\section{Discussion}
This document presents a systematic review of the literature basing on seven (7) articles that point out different security mechanisms to secure containerised based operating systems from MitM attacks. The next sub-sections discuss the evaluation of the collected data-sets and the analysis of the work.
\subsection{Evaluation of data-sets}
Verification of the studied techniques and results are compared with other techniques inline with securing containerised based operating systems. The evaluation of the proposed approaches in the study is done using "real" data as explained below:
\begin{enumerate}
  \item \texttt{"Theoretical approach"} for studies that implement a theoretical mitigation approach in their specified case scenario.
  \item \texttt{"Practical approach"} for studies that implement a practical approach in their specified case scenario.
  \item \texttt{"Machine Learning (ML) / Artificial Intelligence (AI) approach"} for studies that implement a Machine learning approach in their specified case scenario.
\end{enumerate}
Table \ref{taModesti2020Formal} shows the study approaches used as per specified publication author. 

\begin{table}[htbp]
\caption{Real data approaches}
\begin{center}
\resizebox{\columnwidth}{!}{%
\begin{tabular}{|p{0.35\linewidth} | p{0.2\linewidth}| p{0.2\linewidth} | p{0.25\linewidth}|}
\hline
\multicolumn{4}{|c|}{\textbf{Real data}} \\\hline
 \textbf{Publication Author} & \textbf{Theoretical approach} & \textbf{Practical approach}  &\textbf{Machine Learning approach} \\\hline
 \textbf{Mailewa et al \cite{DissanayakaMGK2020vulnerability}} & \checkmark &  &  \\\hline
 \textbf{Ahamed, Zavarsky and Swar \cite{ahamed2021security}} & \checkmark &  &   \\\hline
 \textbf{Wenhao and Zheng \cite{wenhao2020vulnerability}} & \checkmark &  &   \\\hline
 \textbf{Sultan, Ahmad and Dimitriou \cite{SultanAD2019Container}} & \checkmark &  &   \\\hline
 \textbf{Pearson and Plante \cite{pearson2020secure}} & \checkmark & \checkmark  &   \\\hline
 \textbf{Tomar \emph{et al} \cite{tomar2020docker}} & \checkmark & \checkmark  &   \\\hline
 \textbf{Zhang \emph{et al} \cite{zhang2019cm}} & \checkmark & \checkmark  &   \\\hline
\end{tabular}}
\label{taModesti2020Formal}
\end{center}
\end{table}

\subsection{Analysis of the data-sets}
The analysis of the data-sets is done using the SWOT \emph{("Strengths", "Weaknesses", "Opportunities", "Threats")} analysis tool to disseminate the content to detail \cite{ghazinoory2011swot}.
\textbf{\emph{Strengths:}} - The vulnerability assessment tools in the searched data presented quality results for the specified cases. For instance, Mailewa et al \cite{DissanayakaMGK2020vulnerability} presented the use of OpenVAS, Dagda, Portfider, MongoAudit, nmap and Nessus vulnerability assessment tools to come to conclusive results.\\
\textbf{\emph{Weaknesses:}} - The searched data lacked concurrency security design solutions on containerised based operating systems \cite{PervanK2020Parallel}. For instance, they lacked security designs related to programs that are in a collection of independent processes that eventually run in parallel and with the same outcome. Moreover, they lacked solutions embedded with machine learning algorithms that could create at least a training set for detected and classified vulnerabilities. This could eliminate known and future vulnerabilities having the same features in their running processes\\
\textbf{\emph{Opportunites:}} - The security mechanisms in the searched data-sets demonstrate detection of different cyber-attacks while providing confidentiality and integrity of applications in the containers. But they all ignore the security of the underlying container's image. In this case, the attacker may replace the original container image with an infected image to run undetected. The 2019 McAfee \cite{mcafee2019quarterly} report indicated that "Dockerhub" found several of its images used maliciously by unauthorised agents. This creates an opportunity of building a security mechanism that could detect anomalies in containers. However, Karn et al \cite{karn2020cryptomining} proposed a strategy of used Machine Learning (ML) algorithms for detection and classification Kubernetes pods to determine whether the data flow contains running processes or not. \\
\textbf{\emph{Threats:}}- Security on containerised based operating systems is on a rise as one of the major concerns of the researchers. According to the searched data, the mitigation techniques are focused on the outside of the container application and are ignoring the inside of the container. This increases the attack surface of the containers to more likely attacks such as the MitM attacks. \\

\section{The Threat Model}
\subsection {The "Man-in-The-Middle (MiTM)" attack.}
Haselsteiner and Breitfuß \cite{haselsteiner2006security} derived an issue that the adversary intercepts and modifies the communication between the receiver and the sender, who believe they are conversing securely. An attacker can alter the communications so that they reach her or his intended target.

The recently discovered Kubernetes security vulnerability \emph{"CVE-2021-25737}\cite{mitre2021cve25737}," which allows a user to re-route pod traffic to private networks on a Node, highlights the need for developers to double-check the security of Endpoint Slice IPs. Kubernetes, on the other hand, already blocks endpoint internet protocols (IPs) in the localhost or link-local range. Basically, disclosing an external IP address to access a cluster application is more likely to result in a "Man-in-the-Middle (MitM)" attack.

\begin{figure}[h]
\centering
\includegraphics[width=0.5\textwidth]{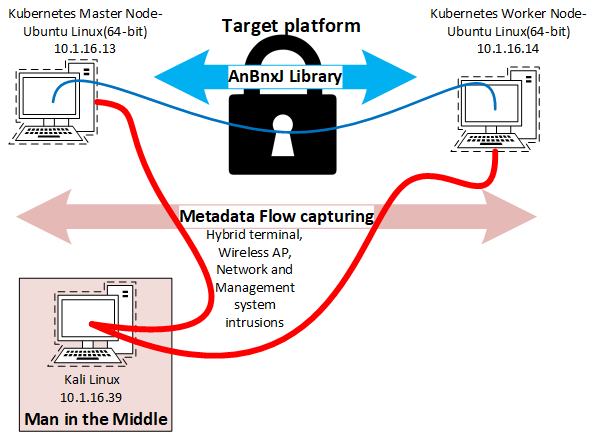}
\caption{\emph{The Attack Tree Scenario}}
\label{fig1}
\end{figure}

Figure \ref{fig1} illustrates the attack tree scenario of the MitM attack and the proposed AnBxJ java security library to provide an API and cryptographic primitives between the master node (10.1.16.13) and the worker node (10.1.16.14). The master-node and worker node are configured in separate containers and on separate linux based operating systems. 
\subsection {Vulnerability Assessment}
\begin{figure}[h]
\centering
\includegraphics[width=0.5\textwidth]{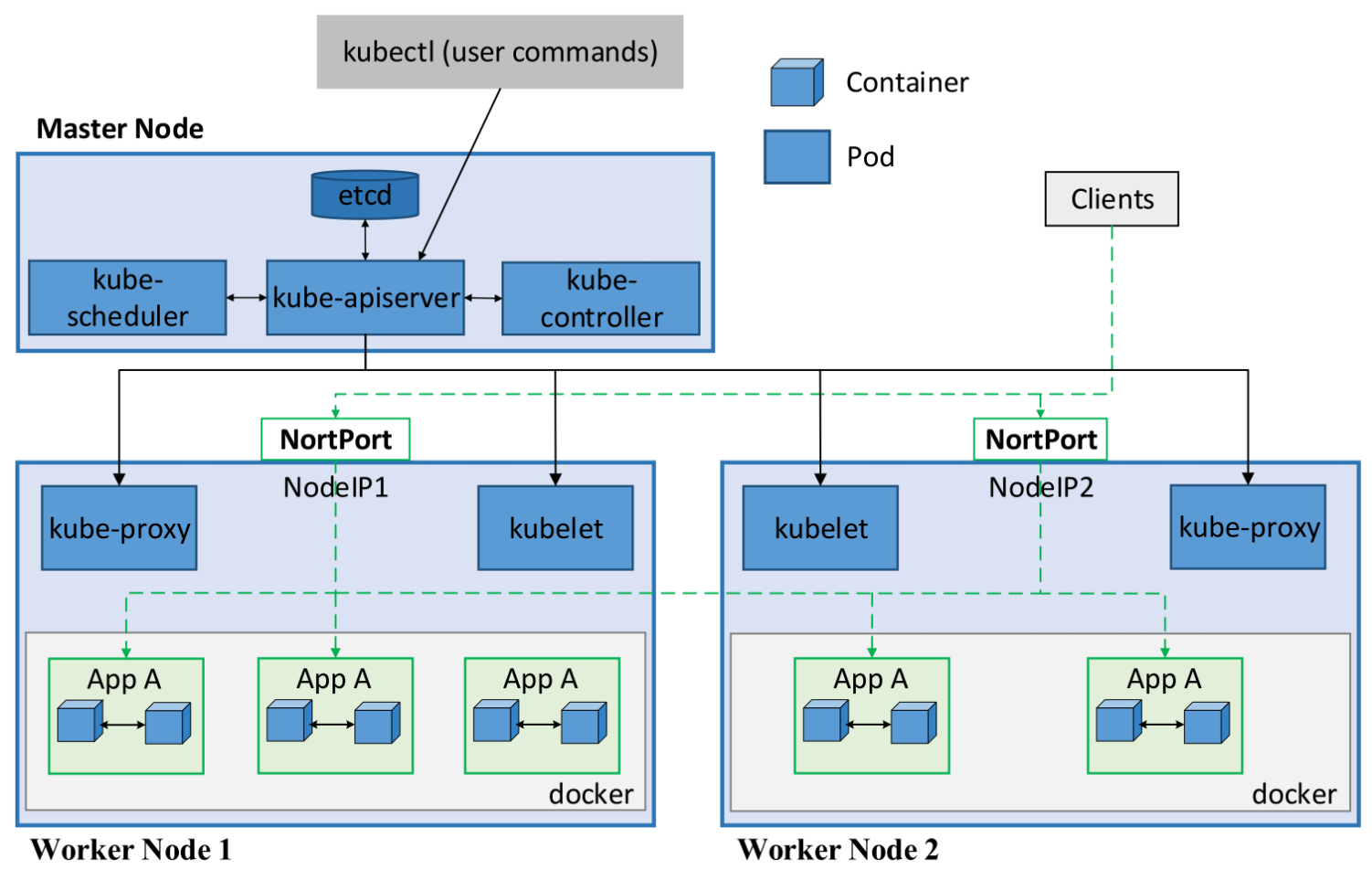}
\caption{\emph{Kubernetes Architecture \cite{kubernetes2021}}}
\label{fig2}
\end{figure}

In order to assess effectively the discussed vulnerability, it is very important to understand the architecture of Kubernetes as illustrated in figure \ref{fig2}. The master node is in charge of controlling and managing the cluster. The \emph{etcd, kube-scheduler, kube-apiserver} and \emph{kube-controller} are the four essential components \cite{kubernetes2021}. The \emph{etcd} is a datastore that is used to store the cluster's configuration and status. The Kubernetes control plane's front end is the \emph{kube-apiserver}. To connect with the Kubernetes cluster, the user or management request must communicate with the \emph{kube-apiserver}. The \emph{kube-scheduler} keeps an eye on unscheduled pods and allocates them to a node based on a variety of parameters, including resource constraints, affinity, and anti-affinity policies. And to keep the cluster in the proper condition, the \emph{kube-controller} keeps an eye on it all the time.

Mailewa \emph{et al} \cite{DissanayakaMGK2020vulnerability} asserts that vulnerability assessment and management is a very important part in the information technology risk management. According to Mailewa \emph{et al} \cite{DissanayakaMGK2020vulnerability}, vulnerability assessment must be prioritised in a likelihood manner ("Low", "Medium", and "High") through an initial analysis of the root cause of the vulnerability in the system's current state. However in this research report, the author considers an initial analysis on the presence of the \emph{"CVE-2021-25737}\cite{mitre2021cve25737} vulnerability on Kubernetes pods.  Moreover, an open-source OpenVAS vulnerability assessment scanner was used to determine the severity of \emph{"CVE-2021-25737} vulnerability \cite{greenbone2021openvas}.

\section{Proposed Security Model and Design}
\begin{figure}[h]
\centering
\includegraphics[width=0.4\textwidth]{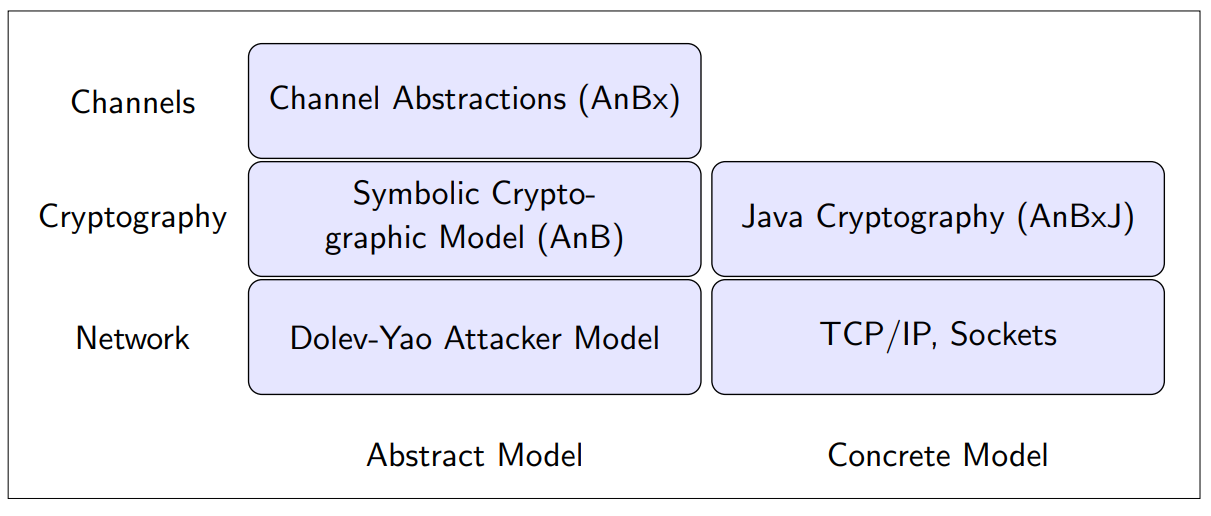}
\caption{\emph{Abstract and Concrete model \cite{Modesti2020Formal}.}}
\label{fig3}
\end{figure}

Security protocols and mechanisms are essential for building secure and stable distributed applications such as Kubernetes, however, their implementation is quite challenging and error prone. Ahamed, Zavarsky and Swar \cite{ahamed2021security} presented a security audit mechanism on Docker containers as well as a cloud architecture embedded with java codes. The author proposed a security mechanism based on a simple language AnB to provide an abstract model and concrete model as shown in figure \ref{fig3} for formal verification and automatic Java implementation \cite{Modesti2020Formal,oracle2021jca}. The proposed architecture is a basis of concurrency java script programs embedded with containerised based programs that are running parallel independent processes, collectively to arrive at the same outcome \cite{PervanK2020Parallel, oracle2021jca}. Figure \ref{fig8} illustrates the proposed secure docker architecture embedded with the AnBxJ security library.

\begin{figure}[h]
\centering
\includegraphics[width=0.4\textwidth]{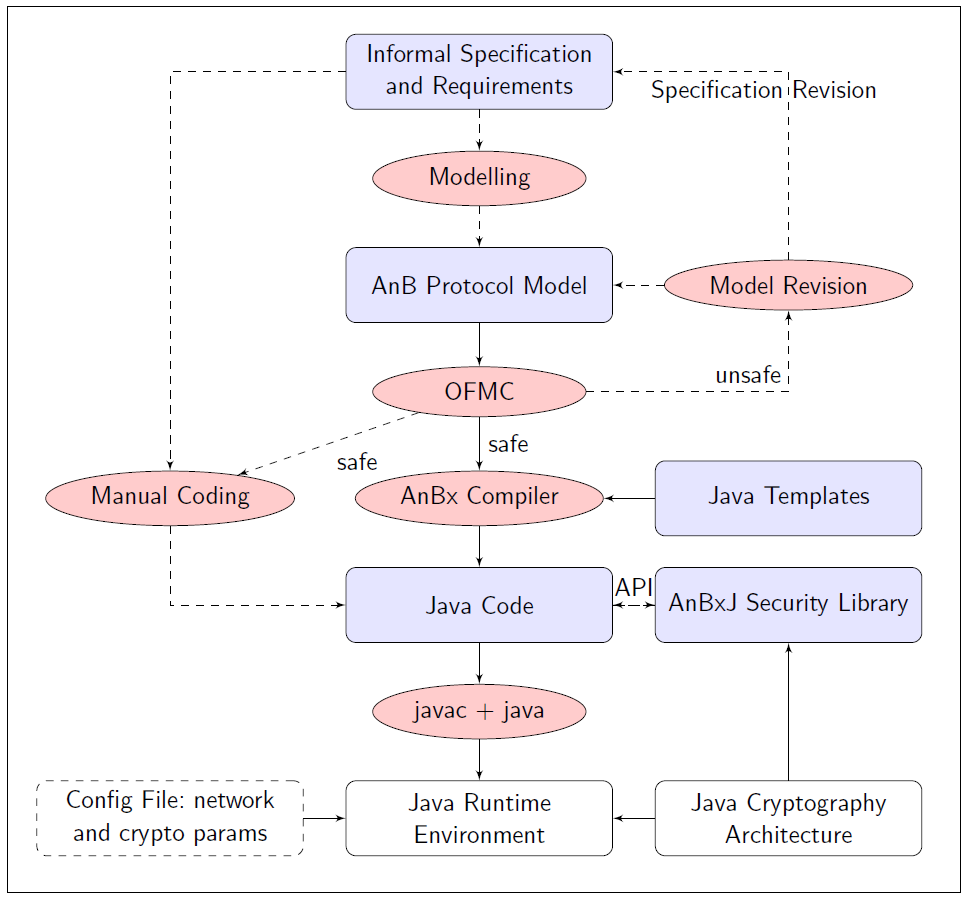}
\caption{\emph{Framework: Workflow and Toolkit \cite{Modesti2020Formal}}}
\label{fig6}
\end{figure}

\begin{figure}[h]
\centering
\includegraphics[width=0.5\textwidth]{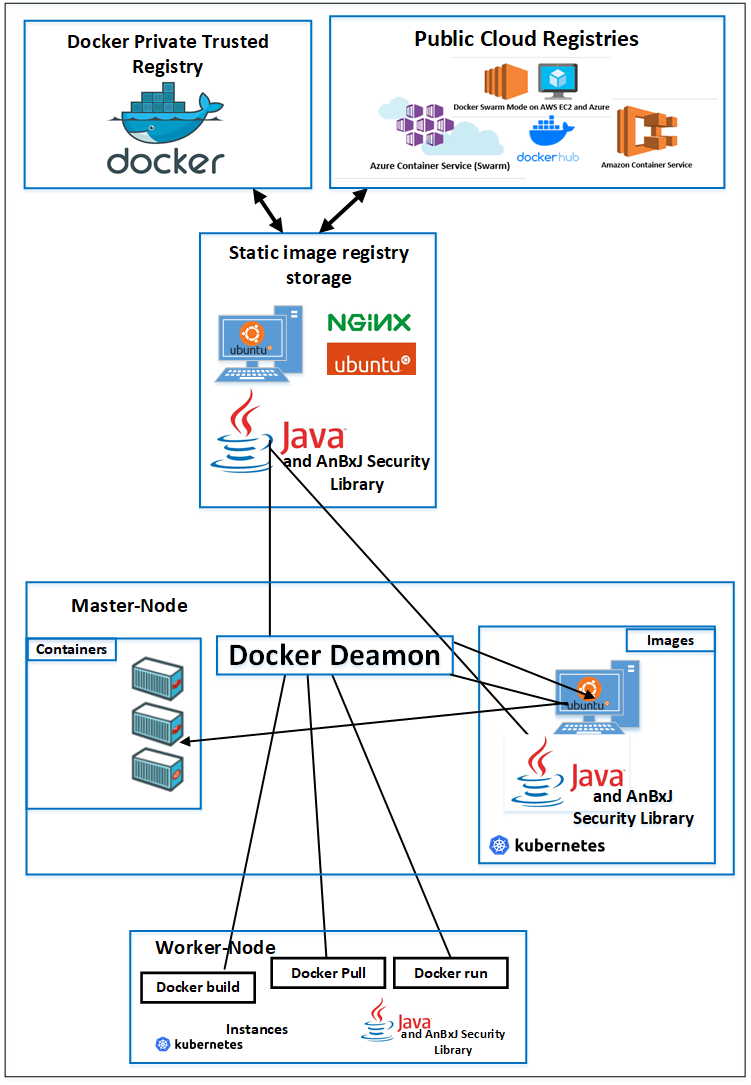}
\caption{\emph{Proposed Secure Docker architecture with AnBxJ security Library.}}
\label{fig8}
\end{figure}

\section{Implementation and Evaluation}
A sample scenario artefact was designed to illustrate the researched effective mechanisms of securing docker containers and Kubernetes pods. Three virtual machines were installed on a vSphere VMware platform. Two of these were Ubuntu 18.04.6 versions and one was Kali Linux. One Ubuntu machine was to serve as the master node and the other Ubuntu machine to serve as the worker node. The master node machine (10.1.16.13) is set to have 2core CPUs and 4GB RAM as well as the worker node (10.1.16.14). The ubuntu machines must go into a series of installation commands for the Kubernetes pods to work effectively\cite{prasath2020setup}. For instance the \emph{swap} has to be disabled on both machines with a command \emph{"swapoff -a"}. The docker 20.10.7 version is installed on both Ubuntu machines. By using the \emph {"sudo systemctl enable docker"} command to enable the installed docker on both machines, activates the docker into the running mode. For Kubernetes to run effectively, it requires the installation of tools \emph{"kubeadm, kubelet and kubectl"} on both the worker node machine and the master node machine \cite{Casalicchio2019Container9,kubernetes2021}.

The AnBxJ library safeguarded communication between the master-node and the worker-node, therefore the application required development methods and testing procedures for recorded meta-data. Facts in modelling security attributes generated during the compilation chain were required for the study of the AnB security goals \cite{Modesti2015AnBx,Modesti2020Formal}.

\subsection {Connection-Queue}
The programmed Java code is designed to accept connections during authentication processes, then close connections in cases of unauthorised access and at the end of each session. Both the worker-node and master-node received private keys as a result of the encryption operations.
\subsection {Multi-listener}
\begin{figure}[h]
\centering
\includegraphics[width=0.5\textwidth]{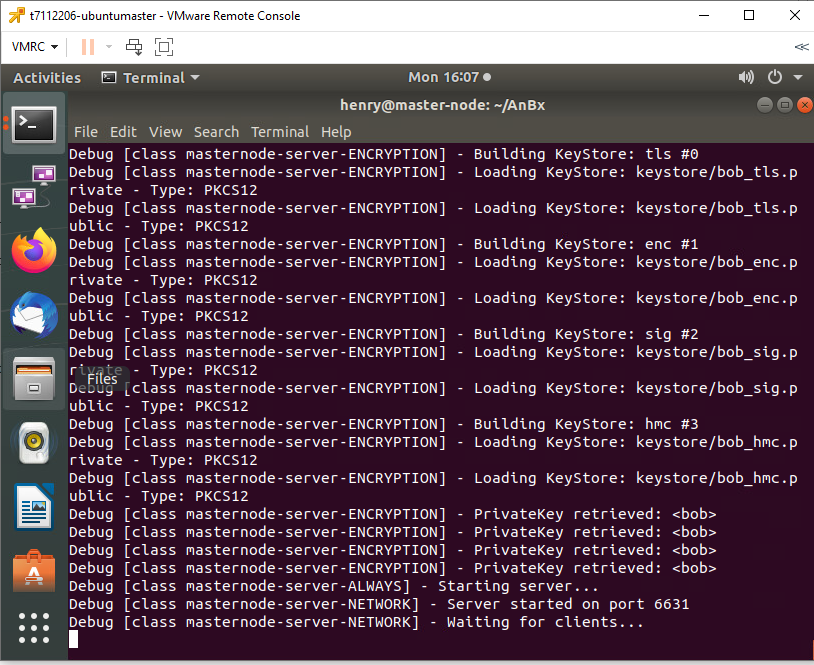}
\caption{\emph{Master-node waiting for clients.}}
\label{fig7}
\end{figure}

A set of compiled java codes activated a multi-listener during the session on the master-node server. The master-node server had the capacity of listening to only connections configured in the \textbf{\emph{".PROPERTIES" file}}.
Figure \ref{fig7} illustrates the master-node server waiting for connections as configured in the \textbf{\emph{".PROPERTIES" file}}  to come via its safe and resilient channel (In this case; Port 6631).

\subsection {Detection}
A critical examination was done on the metadata flow for endpoint addresses under the list of EndpointSlices to see if the \emph{"CVE-2021-25737}\cite{mitre2021cve25737} vulnerability was exploited. Using the openVAS vulnerability assessment scanner, the severity of MitM attacks was shown as high on the list of EndpointSlices \cite{greenbone2021openvas}.  
\subsection {Mitigation}
A Study \cite{terzolo2021enabling} shows a mitigation mechanism through construction of a validating admission webhook that prohibits EndpointSlices with endpoint addresses from exploiting this vulnerability (\emph{"CVE-2021-25737}) without having to upgrade kube-apiserver. However, a policy establishment is required to impose this limitation in case of an existing admission policy mechanism as well as the requirement of upgrading the kube-episever.

\begin{figure}
 \centering 
 \includegraphics[width=0.5\textwidth]{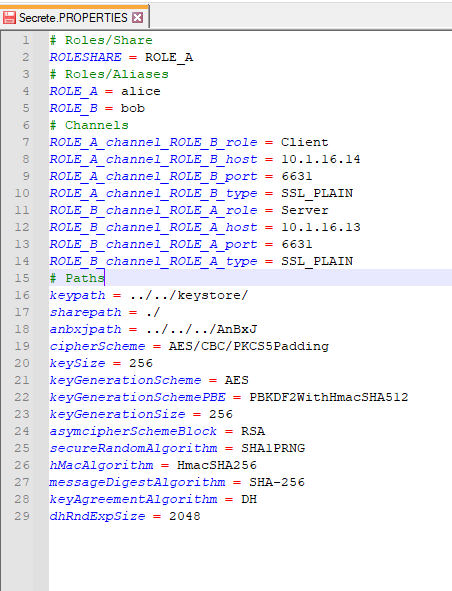}
 \caption{\emph{sample properties file}}
 \label{fig4}
\end{figure}

\begin{figure}
 \centering 
 \includegraphics[width=0.5\textwidth]{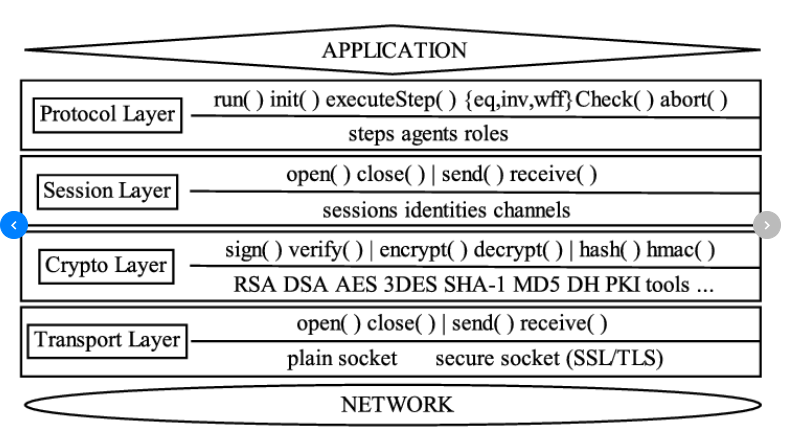}
 \caption{\emph{AnBxJ Java library architecture \cite{Modesti2020Formal}}}
 \label{fig5}
\end{figure}

In this regard, the author proposed a security mechanism that could eliminate all the hardships of timely upgrading of the kube-apiserver and easily configured by software developers. The security mechanism was designed basing on the Java programming language and cryptographic primitives offered by the Java Cryptography Architecture (JCA) \cite{oracle2021jca}. Moreover, the multi-listener and connection-queue feature properties in the proposed mechanism, give an added advantage to eliminate malicious connections proactively. Thus, eliminating the EndpointSlices with endpoint addresses from exploiting the \emph{"CVE-2021-25737} vulnerability. The author used the AnBxJ library which is a sub section of an open-source tool called the AnBx Compiler and Code Generator \cite{Modesti2015AnBx}. The \textbf{\emph{".PROPERTIES" file}} as shown in figure \ref{fig4} and several Java script codes to be found in the Github \cite{kabuye2021exploiting} platform. These codes and the AnBxJ Java library were compiled by the AnBx Compiler and Code Generator \cite{Modesti2015AnBx}. Figure \ref{fig5} illustrates the AnBxJ Java library architecture.

\section{Future work and Conclusion}
Java as a programming language has been ignored by many security developers through bias and the sensitivity of cases when writing their scripts for debugging. The researched security mechanisms in this work shows that the power of security in API applications remains in java cryptographic architectures (JCA). Moreover, this paper presents a systematic review of reviewed studies on securing containerised-based operating systems and demonstrates a java security mechanism. The java security mechanism proposed in this review, is based on a simple language AnB (Alice and Bob) notation to provide an abstract and concrete model. Further studies may focus on java security firewalls on concurrency designs using the AnB novel notation and machine learning (ML) algorithms to detect and classify anomalies in containerised based operating systems.
\section{Ethical statement}
The project is a research-based investigation that used Teesside University resources to study various digital platforms online while adhering to Teesside University's ethics and behaviour. The artefact requires no additional human interaction and was created using the University's vSphere VMware virtualised platform, which Teesside University has fully protected to prevent any harm to any organisation or individual.

\bibliographystyle{IEEEtran}
\bibliography{references}

\end{document}